\documentclass[conference,10pt]{IEEEtran}
\usepackage[T1]{fontenc}

\usepackage{amssymb,amsmath,amsfonts,amsthm}
\usepackage{mathrsfs}
\usepackage{cite}
\usepackage{array}
\usepackage{tabularx}
\usepackage[table]{xcolor}
\usepackage{multicol, multirow}
\usepackage{color}
\usepackage{mathtools}
\usepackage{subcaption}
\usepackage{mathtools}
\usepackage{tikz,pgf,circuitikz}
\usetikzlibrary{calc,positioning,mindmap,trees,decorations.pathreplacing}
\usepackage{graphicx}
\usepackage{bbm}
\usepackage[utf8]{inputenc}
\usepackage{algorithm}
\usepackage{algpseudocode}

\renewcommand{\IEEEQED}{\IEEEQEDopen}

\usepackage[colorinlistoftodos,prependcaption,backgroundcolor=black!5!white,bordercolor=red]{todonotes}

\IEEEoverridecommandlockouts
\ifCLASSINFOpdf
\else
\fi
\usepackage[cmintegrals]{newtxmath}
\begin{document}

\title{Over-the-Air Federated Learning with Phase Noise: Analysis and Countermeasures\\
\thanks{This work was supported in part by ELLIIT, the Swedish Research Council (VR), and
	the Knut and Alice Wallenberg (KAW) Foundation.}
}

\author{\IEEEauthorblockN{Martin Dahl}
\IEEEauthorblockA{\textit{Dept. of Electrical Engineering (ISY)} \\
\textit{Linköping University, Sweden}\\
martin.dahl@liu.se}
\and
\IEEEauthorblockN{Erik G. Larsson}
\IEEEauthorblockA{\textit{Dept. of Electrical Engineering (ISY)} \\
\textit{Linköping University, Sweden}\\
erik.g.larsson@liu.se}
}

\maketitle

\begin{abstract}
Wirelessly connected devices can collaborately train a machine learning model using federated learning, where the aggregation of model updates occurs using over-the-air computation. Carrier frequency offset caused by imprecise clocks in devices will cause the phase of the over-the-air channel to drift randomly, such that late symbols in a coherence block are transmitted with lower quality than early symbols. To mitigate the effect of degrading symbol quality, we propose a scheme where one of the permutations Roll, Flip and Sort are applied on gradients before transmission. Through simulations we show that the permutations can both improve and degrade learning performance. Furthermore, we derive the expectation and variance of the gradient estimate, which is shown to grow exponentially with the number of symbols in a coherence block.

\end{abstract}


\section{Introduction}
With an increasing number of wirelessly connected devices the question arises how to collaboratively train machine learning (ML) models with device data. Meanwhile it is important to consider user privacy and communication resource constraints. A promising method of distributed ML is federated learning (FL) \cite{mcmahan2017communication} where devices aggregate model updates instead of sharing data, giving partial privacy protection \cite{zhu2019deep} and relief from transmitting large volumes of data. In recent works the application of wireless FL is explored \cite{gafni2022federated}, demonstrating important aspects such as power control for energy efficiency, resource allocation as well as compression of model updates through sparsification and quantization. 

One method for wireless FL is over-the-air computation (OAC), exploiting the superposition nature of radio waves for aggregation of model updates sent with analog modulation \cite{csahin2023survey}. In general, OAC can be used to compute any nomographic function \cite{goldenbaum2014nomographic} for a wider range of applications than FL, such as control and sensing \cite{park2021optimized, shao2022bayesian}. As communication and computation are merged, the main benefit of OAC is the $\mathcal{O}(1)$ aggregation resource use instead of $\mathcal{O}(K)$ from sequential communication and computation with $K$ devices.

OAC schemes are either coherent or non-coherent, depending on the availability of channel state information (CSI) at the devices and base station. While schemes requiring no or partial CSI have been studied \cite{goldenbaum2013robust, csahin2023distributed, tegin2023federated}, we consider a scenario where estimation of CSI is required at transmitters to enable channel inversion, similar to \cite{zhu2019broadband}. In OAC devices must also perform efficient power control to counteract fading \cite{cao2020optimized} and coarse time-synchronization with other devices on a frame level \cite{goldenbaum2013robust}. Previous works have studied the effect of inaccurate CSI \cite{chen2022over}, inaccurate phase \cite{sery2020analog} as well as carrier frequency offset (CFO) \cite{you2023broadband, shao2021federated,guo2021over} caused by inaccurate local oscillators in devices. While solutions such as GPS and wired synchronization exist to remove CFO, these are typically expensive and impractical in the wireless context. Cheaper wireless protocols such as \cite{abari2015airshare} have been studied but will not remove CFO completely, therefore some degree of CFO is always expected to be present. 

In \cite{you2023broadband, shao2021federated,guo2021over} the CFO of devices was assumed fixed, which causes a linear drift of phase. However, in a practical system the device oscillators are noisy and drift over time, resulting in time-varying phase noise as compared to linear phase drift. Phase noise is commonly modeled as a Wiener process, a random walk with Gaussian increments, which we adopt herein \cite{demir1998phase, tomba1998effect, petrovic2007effects}. The variance of Wiener phase noise increases linearly with time and causes a random rotation of the transmitted symbols that gets worse over time. Eventually the oscillator phase gets completely out of sync and must be re-aligned using pilot signaling or calibration measurements.

As transmitted symbols carry gradient elements, Wiener phase noise implies that gradient elements transmitted early are received more accurately than those transmitted later. In relation to this, we note that the importance of specific gradient elements for ML-models are in general not equal, something observed in the context of gradient sparsification \cite{alistarh2018convergence}. Moreover, gradient elements belonging to the first layers of a neural network could be more important than gradient elements of the final layers \cite{zhang2022all, ko2022not}. The question then arises as to how serious the effect of the phase noise is on transmitted gradients and what countermeasures can be applied.

\textbf{Contributions:} We study the effect of phase noise caused by noisy device oscillators on over-the-air federated learning. We propose a scheme where permutations are applied to gradients before transmission which changes the order in which specific gradient elements are transmitted. This enables prioritizing important gradient elements by transmitting them with earlier symbols, which are in turn received more accurately. Furthermore, we derive the expectation and variance of the gradient estimate at the base station and demonstrate by simulation that the effect of permutation on learning performance can be significant. 



\section{Problem Formulation and System Model}
\subsection{Distributed Optimization}
We consider a set of distributed devices indexed by $k\in\{0,1,...,K-1\}$ with local parameters $\boldsymbol{\theta}_k\in\mathbb{R}^D$. Each device holds a local objective function $f_k(\boldsymbol{\theta}_k)$ and dataset $\mathcal{D}_k\subset \mathbf{R}^{C\times D}$ where $D$ is even for ease of notation. The local objective follows:
\begin{equation}
	f_k(\boldsymbol{\theta}) = \frac{1}{C}\sum_{c=1}^{C}\mathcal{L}(\delta_{k,c}, \boldsymbol{\theta}) \equiv \mathcal{L}(\mathcal{D}_k, \boldsymbol{\theta}),
\end{equation}
which can be considered the average loss of a machine learning model with some loss function $\mathcal{L}$ over all samples $\delta_{k,c}\in\mathcal{D}_k$. Finally, the goal is to optimize the global objective:
\begin{equation}\label{eq:objective}
	\underset{\boldsymbol{\theta}}{\text{ min }}F(\boldsymbol{\theta}) = \frac{1}{K}\sum_{k=0}^{K-1}f_k(\boldsymbol{\theta}).
\end{equation}

\subsection{Federated Learning}
A local optimum of objective (\ref{eq:objective}) can be found using federated learning: In every iteration $n\in\{0,1,...,N-1\}$, every device $k$ computes its local batch-stochastic gradient
\begin{equation}
	\mathbf{g}_k^{(n)} = \left[g_{k,0}^{(n)}, ..., g_{k,D-1}^{(n)}\right]^\text{T} = \frac{1}{B}\nabla \mathcal{L}\left(\mathcal{B}_k^{(n)}, \boldsymbol{\theta}^{(n)}\right),
\end{equation}
with randomly sampled batch $\mathcal{B}_k^{(n)} \subset \mathcal{D}_k$, $|\mathcal{B}_k^{(n)}| = B\in\mathbb{N} \text{ } \forall n$, $\boldsymbol{\theta}_k^{(n)} = \boldsymbol{\theta}^{(n)}$. Then all $\mathbf{g}_k^{(n)}$are transmitted to the base station and aggregated into $\mathbf{g}^{(n)}$ as follows:
\begin{equation}\label{aggregated_g}
	\mathbf{g}^{(n)} = \frac{1}{K}\sum_{k=0}^{K-1}\mathbf{g}_k^{(n)},
\end{equation}
Next, the base station updates the global model using step-size $\gamma$:
\begin{equation}
	\boldsymbol{\theta}^{(n+1)} = \boldsymbol{\theta}^{(n)} - \gamma \mathbf{g}^{(n)}. 
\end{equation}
Finally $\boldsymbol{\theta}^{(n+1)}$ is transmitted to all devices such that $\boldsymbol{\theta}_k^{(n+1)} = \boldsymbol{\theta}^{(n+1)}$ and the next iteration $n+1$ proceeds until $N-1$.

\subsection{System Model}
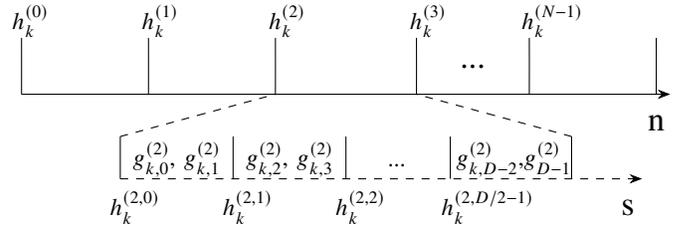
\begin{figure}[!t]
	\begin{center}
		\begin{circuitikz}[scale=0.75, transform shape]
			\draw [->, >=Stealth] (8.75,14) .. controls (14.75,14) and (14.75,14) .. (20.25,14);
			\draw [](8.75,14) to[short] (8.75,15);
			\draw [](11,14) to[short] (11,15);
			\draw [](13.25,14) to[short] (13.25,15);
			\draw [](15.75,14) to[short] (15.75,15);
			\draw [](17.75,14) to[short] (17.75,15);
			\node [font=\LARGE] at (16.75,14.5) {...};
			\node [font=\LARGE] at (20,13.5) {n};
			\draw [->, >=Stealth, dashed] (10.5,12.5) .. controls (15,12.5) and (15,12.5) .. (19.75,12.5);
			\node [font=\LARGE] at (19.5,12) {s};
			\draw [short] (10.5,13.25) .. controls (10.5,13) and (10.5,13) .. (10.5,12.5);
			\draw [short] (18.5,13.25) .. controls (18.5,13) and (18.5,13) .. (18.5,12.5);
			\draw [short] (12.5,13.25) .. controls (12.5,13) and (12.5,13) .. (12.5,12.5);
			\draw [short] (14.5,13.25) .. controls (14.5,13) and (14.5,13) .. (14.5,12.5);
			\draw [short] (16.35,13.25) .. controls (16.35,13) and (16.35,13) .. (16.35,12.5);
			\node [font=\large] at (10.75,12) {$h_k^{(2,0)}$};
			\node [font=\large] at (12.75,12) {$h_k^{(2,1)}$};
			\node [font=\large] at (14.75,12) {$h_k^{(2,2)}$};
			\node [font=\large] at (17.0,12) {$h_k^{(2,D/2-1)}$};
			\node [font=\large] at (8.9,15.3) {$h_k^{(0)}$};
			\node [font=\large] at (15.4,12.75) {...};
			\node [font=\large] at (11.2,15.3) {$h_k^{(1)}$};
			\node [font=\large] at (13.45,15.3) {$h_k^{(2)}$};
			\node [font=\large] at (15.95,15.3) {$h_k^{(3)}$};
			\node [font=\large] at (18.15,15.3) {$h_k^{(N-1)}$};
			\draw [](20,14) to[short] (20,15);
			\node [font=\large] at (11.5,12.85) {$g_{k,0}^{(2)}$, $g_{k,1}^{(2)}$};
			\node [font=\large] at (13.5,12.85) {$g_{k,2}^{(2)}$, $g_{k,3}^{(2)}$};
			\node [font=\large] at (17.475,12.85) {$g_{k,D-2}^{(2)}$,$g_{D-1}^{(2)}$};
			\draw [dashed] (10.5,13.25) to[short] (13.25,14);
			\draw [dashed] (18.5,13.25) to[short] (15.75,14);
		\end{circuitikz}
	\end{center}
	\caption{Symbols carrying gradient elements in each coherence block for device $k$.}
	\label{schematic}
\end{figure}

We assume Rayleigh block fading, with $h_k$ being the CSI of the channel from device $k$ to the base station at time $Tn$: 

\begin{equation}
	\begin{split}
		&h_k(Tn) \equiv h_k^{(n)} = \sqrt{\beta}_k^{(n)}e^{j\phi_k^{(n)}} \sim \mathcal{CN}\left(0, \sigma^2_h\right), \text{i.i.d}.
	\end{split}	
\end{equation}
with coherence-time $T$ such that the symbol time $\tau = \frac{2T}{D}$. We model the phase noise of $\phi_k$ within coherence blocks as a Wiener process 

\begin{equation}
	h_k(Tn + s\tau) \equiv h_k^{(n,s)} = \sqrt{\beta}_k^{(n)}e^{j\phi_k^{(n,s)}},
\end{equation}
where
\begin{equation}
	\begin{split}
		&\phi_k^{(n,s)} = \phi_k^{(n,s-1)} + e_k^{(n,s)},\\
		& \phi_k^{(n,0)} = \phi_k^{(n)},\\
		& e_k^{(n,s)}\sim\mathcal{N}\left(0,\sigma^2_e\right), \text{ i.i.d. and } e_k^{(n,0)}=0,\\
		& s\in \{0,1,...,D/2-1\}.
	\end{split}
\end{equation}
At the start of each coherence block the devices perfectly estimate $h_k^{(n)}$. Then each transmitted symbol, as demonstrated in Figure \ref{schematic}, can be written as 

\begin{equation}
	x_k^{(n,s)} =
	\begin{cases}
		\frac{g_{k,2s}^{(n)} + jg_{k,2s+1}^{(n)}}{h_k^{(n)}} &\left|h_k^{(n)}\right|^2 \geq t \\
		0& \text{else},
	\end{cases}
\end{equation}
and the received symbol can be written as
\begin{equation}
	\begin{split}
		&y\left(Tn + s\tau\right) \equiv y^{(n,s)} = \sum_{k=0}^{K-1}h_k^{(n,s)}x_{k}^{(n,s)} + w^{(n,s)},
	\end{split}	
\end{equation}
with thermal noise
\begin{equation*}
	w^{(n,s)} \sim \mathcal{CN}\left(0, \sigma^2_w\right).
\end{equation*}
 We apply truncated channel inversion \cite{zhu2019broadband} with threshold $t\in\mathbb{R}$, assuming $\left|g_{k,2s}^{(n)} + jg_{k,2s+1}^{(n)}\right|$ always sufficiently small to satisfy power constraint $\left|x_k^{(n,s)}\right|^2 \leq P\in\mathbb{R}$. This gives 
\begin{equation}
	\begin{split}
		&y^{(n,s)} = \sum_{k=0}^{K-1}e^{-j\sum_{i=0}^{s}e_k^{(n,i)}}\left(g_{k,2s}^{(n)} + jg_{k,2s+1}^{(n)}\right)\mathbb{I}_k^{(n)} + w^{(n,s)},
	\end{split}	
\end{equation}
where $\mathbb{I}_k^{(n)} = 1$ if $\left|h_k^{(n)}\right|^2 \geq t$, else $0$.
\section{Estimate of Aggregated Gradient}
The estimated aggregated gradient at the base station is $\Hat{\mathbf{g}}^{(n)}$ where
\begin{equation}\label{estimate}
	\Hat{g}^{(n)}_d = \frac{a_d}{K}
	\begin{cases}
		\Re\left(y^{\left(n,\frac{d}{2}\right)}\right) & d \text{ even}\\
		\Im\left(y^{\left(n,\frac{d-1}{2}\right)}\right) & d \text{ odd},
	\end{cases}
\end{equation}
with normalizing factor $a_d\in\mathbb{C}$.

\subsection{Expectation and Variance of Estimate}
Given $\mathbf{g}^{(n)}$, for even $d$
\begin{equation}\label{eq:firstmoment}
	\begin{split}
		&\mathbb{E}\left[\Hat{g}^{(n)}_d\right] = \mathbb{E}\left[\Re\left(\frac{a_d}{K}\sum_{k=1}^{K-1}e^{-j\sum_{i=0}^{d/2}e_k^{(n,i)}}\left(g_{k,d}^{(n)} + jg_{k,d+1}^{(n)}\right)\mathbb{I}_k^{(n)}\right)\right]\\   
		 &= \Re\left(\frac{a_d}{K}\sum_{k=0}^{K-1}\mathbb{E}\left[e^{-j\sum_{i=0}^{d/2}e_k^{(n,i)}}\left(g_{k,d}^{(n)} + jg_{k,d+1}^{(n)}\right)\mathbb{I}_k^{(n)}\right]\right)\\
		 &= \Re\left(\frac{a_d}{K}\sum_{k=0}^{K-1}\text{exp}\left(\frac{-d\sigma^2_e}{4}\right)\left(g_{k,d}^{(n)} + jg_{k,d+1}^{(n)}\right)Q_{\mathcal{X}^2_2}\left(\frac{2t}{\sigma_h^2}\right)\right)
		 \\&= \frac{a_d\text{exp}\left(\frac{-t}{\sigma^2_h}-\frac{d\sigma^2_e}{4}\right)}{K}\sum_{k=0}^{K-1}g_{k,d}^{(n)},
	\end{split}
\end{equation}
where
\begin{equation}\label{eq:constants_1}
	\begin{split}
	&Q_{\mathcal{X}^2_2}\left(\frac{2t}{\sigma_h^2}\right) \equiv P\left(\left|h_k^{(n)}\right|^2 \geq t\right) = \text{exp}\left(\frac{-t}{\sigma^2_h}\right),
	\end{split}
\end{equation}
giving an unbiased estimate of (\ref{aggregated_g}) for $a_d = \text{exp}\left(\frac{t}{\sigma^2_h} + \frac{d\sigma^2_e}{4}\right)$. For odd $d$ the derivation is similar and $a_d = \text{exp}\left(\frac{t}{\sigma^2_h} + \frac{(d-1)\sigma^2_e}{4}\right)$. Furthermore, for even $d$ the variance of the estimator is given by
\begin{equation}\label{variance}
	\begin{split}
		&\text{Var}(\hat{g}_d^{(n)})= \frac{\text{exp}\left(\frac{t}{\sigma^2_h}+\frac{d\sigma^2_e}{2}\right)}{2K^2}\dots\\
		&\sum_{k=0}^{K-1}\left(\bigg(1 + \text{exp}\left(-d\sigma^2_e\right)-2\text{exp}\left(\frac{-t}{\sigma^2_h}-\frac{d\sigma^2_e}{2}\right)\right)g_{k,d}^2\\ 
		&+ \bigg(1 - \text{exp}\left(-d\sigma^2_e\right)\bigg)g_{k,d+1}^2\bigg)+ \frac{a_d^2\sigma^2_w}{2K^2},
	\end{split}
\end{equation}
where for uneven $d$, $d$ in (\ref{variance}) is exchanged with $d-1$.
\begin{proof}
	See Appendix \ref{Appendix}.
\end{proof}
\section{Gradient Permutation}
Since the effect of phase noise will not be equal for all symbols, we propose a scheme where every device applies a permutation $\tilde{\mathbf{g}}_k^{(n)} \equiv \mathbf{P}^{(n)}\mathbf{g}_k^{(n)}$ before transmission. The base station applies the reverse permutation on the estimated permuted gradient $\hat{\tilde{\mathbf{g}}}^{(n)}$ to obtain $\hat{\mathbf{g}}^{(n)}\equiv \left(\mathbf{P}^{(n)}\right)^{-1}\hat{\tilde{\mathbf{g}}}^{(n)}$. By the "original" permutation we refer to $\mathbf{P} = \mathbf{I}$, where the input layers come first and output layers last.
\subsection{Flip}
The order of the gradient elements is flipped
\begin{equation}
	\tilde{g}_{k,d}^{(n)} = g_{k,D-1-d}^{(n)},
\end{equation}
such the gradient elements of the final layers are transmitted first.

\subsection{Roll}
Let $t = n\cdot D/2 + s + 1$ be the total number of transmissions and $r = (t-1) \% (D/2)$ the roll length where $\%$ is the modulo operator. The roll permutation is 
\begin{equation}
	\tilde{g}_{k,d}^{(n)} = \begin{cases}
		g_{k,d-r}^{(n)}&\text{if } 0 \leq d-r\\
		g_{k,D/2-r}^{(n)}&\text{else},
	\end{cases}
\end{equation}
such that over time every gradient element is repeatedly sent with every degree of phase noise.

\subsection{Sort (Genie Aided)}
At the start of each epoch, every device computes $\mathbf{g}_k^{(n)}$ using one batch. Then $\left|g_{k,d}^{(n)}\right|\text{ }\forall d$ is transmitted over an error-free channel to the base station. First, the base station computes the average of the absolute gradient elements 
\begin{equation}
	\left|g_d^{(n)}\right| \equiv \frac{1}{K}\sum_{k=0}^{K-1}\left|g_{k,d}^{(n)}\right|.
\end{equation}
Next, $\left|g_d^{(n)}\right|$ are sorted from max to min, giving sorted indices

\begin{equation}
	 \mathbf{i} = \text{sort}\left(\left[\left|g_{0}^{(n)}\right|, \left|g_{1}^{(n)}\right|, ..., \left|g_{D-1}^{(n)}\right|\right]\right) \in \mathbb{R}^{D},
\end{equation}
and the sorted permutation
\begin{equation}
	\tilde{g}_{k,d}^{(n)} = g_{k,i_d}^{(n)},
\end{equation}
which is distributed to the devices over an error-free channel. The Sort permutation prioritizes gradient elements of high absolute value, since the high value implies that they will have a high impact on the loss.
\section{Simulations}
The proposed scheme with permutations is tested in an image classification setting, where a convolutional neural network (CNN) \cite{Goodfellow-et-al-2016}, defined in Table \ref{table:1}, is trained with cross-entropy loss on the MNIST digits dataset \cite{lecun1998mnist} with a heterogeneous distribution of data across devices.

\subsection{Learning Hyperparameters and Heterogeneous Data}
The MNIST train dataset consists of 60000 samples representing 10 digit classes (0-9). We sort the train dataset by digit label and split it into 20 shards of size 3000. The network consists of 10 devices that are randomly allocated 2 distinct shards. This gives each device 6000 samples representing between 1 and 4 unique digits, which is a heterogeneous distribution of the MNIST train set. For every device, batch-size is set to $B=5$ giving 1200 batches per epoch, step-size is set to $\gamma=0.01$ and kept constant throughout the learning process. Finally, the MNIST test dataset has 10000 samples with an even distribution of all digits.

\begin{table}[t!]
	\centering
	\begin{tabular}{||c c ||} 
		\hline
		\textbf{Layer} & \textbf{Hyperparameters} \\ [0.5ex] 
		\hline\hline
		Conv2D & In channels=1, Out channels=6, Kernel size=5 \\ 
		MaxPool2D & Kernel size=2, Stride=2 \\
		Conv2D & In channels=6, Out channels=2, Kernel size=5 \\
		Linear & Input size=32, Output size=30 \\
		Linear & Input size=30, Output size=10 \\ [1ex] 
		\hline
	\end{tabular}
	\caption{CNN with 1738 parameters, ReLU activation is used between all layers and Softmax for prediction output.}
	\label{table:1}
\end{table}

\subsection{System Model Hyperparameters}
\begin{figure}[t!]
	\centering
	\begin{subfigure}[b]{1\columnwidth}
		\centering
		\includegraphics[width=0.9\columnwidth]{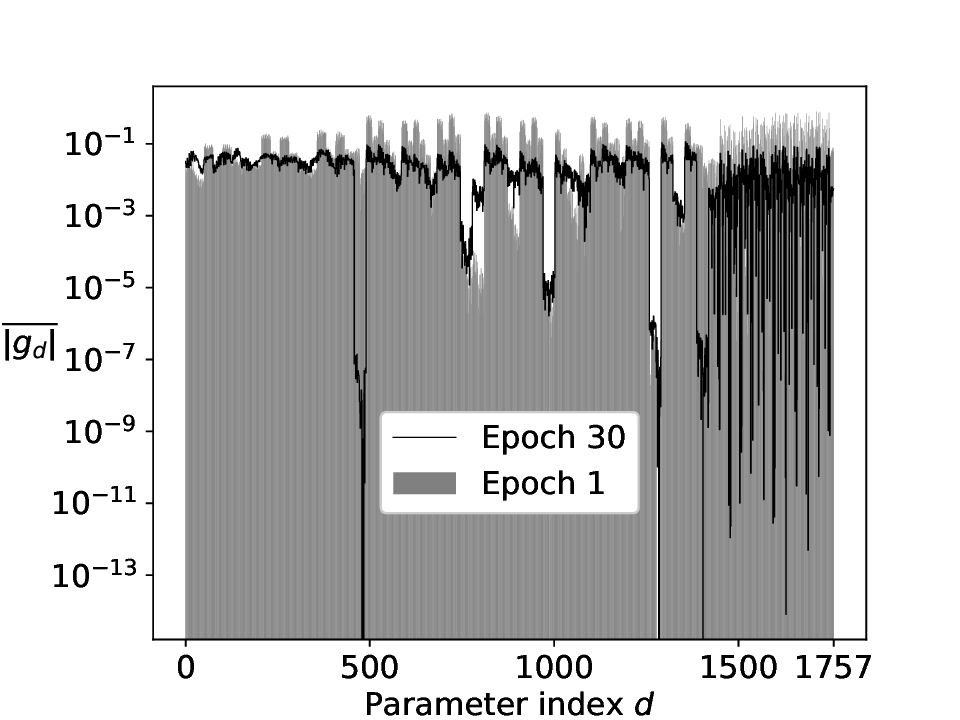}
	\end{subfigure}
	\caption{Average of gradient elements absolute value in epoch 1 and epoch 30. 10 trials. Dips around index 750, 1000, 1250 caused by ReLU activation.}
	\label{fig:abs_grad_1_30}
\end{figure}
\begin{figure}[t!]
\centering
\begin{subfigure}[b]{1\columnwidth}
	\centering
	\includegraphics[width=0.9\columnwidth]{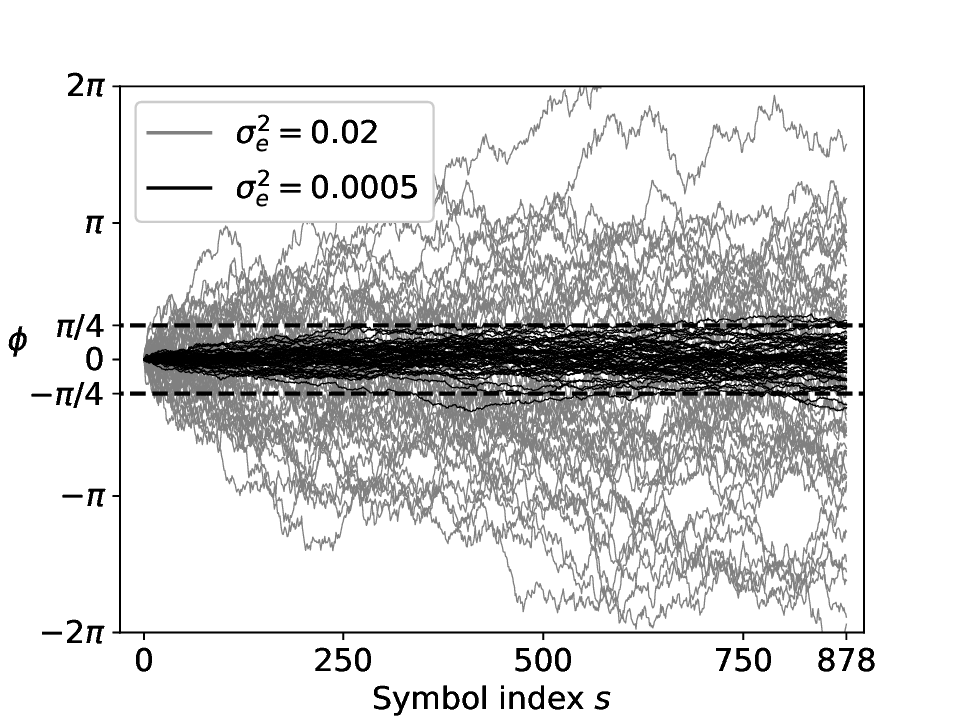}
\end{subfigure}
\caption{50 realizations of low and high phase noise, $\sigma^2_e = 0.0005$ and $\sigma^2_e = 0.02$, respecively.}
\label{fig:phase_drift}
\end{figure}

As seen in Figure \ref{fig:abs_grad_1_30}, most gradient elements apart from some ReLU dips have an absolute value above $10^{-3}$ throughout the training process, so the power of each symbol is approximately $2\cdot10^{-6}$. Thus, for a consistent SNR of approximately $20$ dB we set $\sigma^2_w=2\cdot10^{-8}$. Furthermore, we set $t=0.01$ and $\sigma^2_h=1$ giving $Q_{\mathcal{X}^2_2}\left(\frac{2t}{\sigma_h^2}\right) \approx 0.99$. Phase noise is demonstrated in Figure \ref{fig:phase_drift}, where a low and high phase noise scenario with $\sigma_e^2=0.0005$ and $\sigma_e^2=0.02$ is used, respectively. In the low phase noise scenario all symbol phases are roughly kept within $[-\pi/4, \pi/4]$ which conserves the sign of both the real and complex part of $\mathbf{x}$ when adjacent gradient elements are approximately equal, which is common in over-parameterized neural networks. In the high phase noise scenario this is not the case and error in sign of the reconstructed gradient is more probable. An important note is that because of the coefficient $a_d$ in the estimator (\ref{estimate}) the variance (\ref{variance}) explodes as $d$ grows. For this reason we replace $a_d$ with a more practical $\tilde{a}_d\equiv\text{exp}\left(\frac{t}{\sigma_h^2}\right)$ at the cost of the estimator having a statistical bias. The effect of $\tilde{a}_d$ is equivalent to a scaled step-size.

\subsection{Simulation Results}
\begin{figure}[t!]
	\centering
	\begin{subfigure}[b]{1\columnwidth}
		\centering
		\includegraphics[width=0.9\columnwidth]{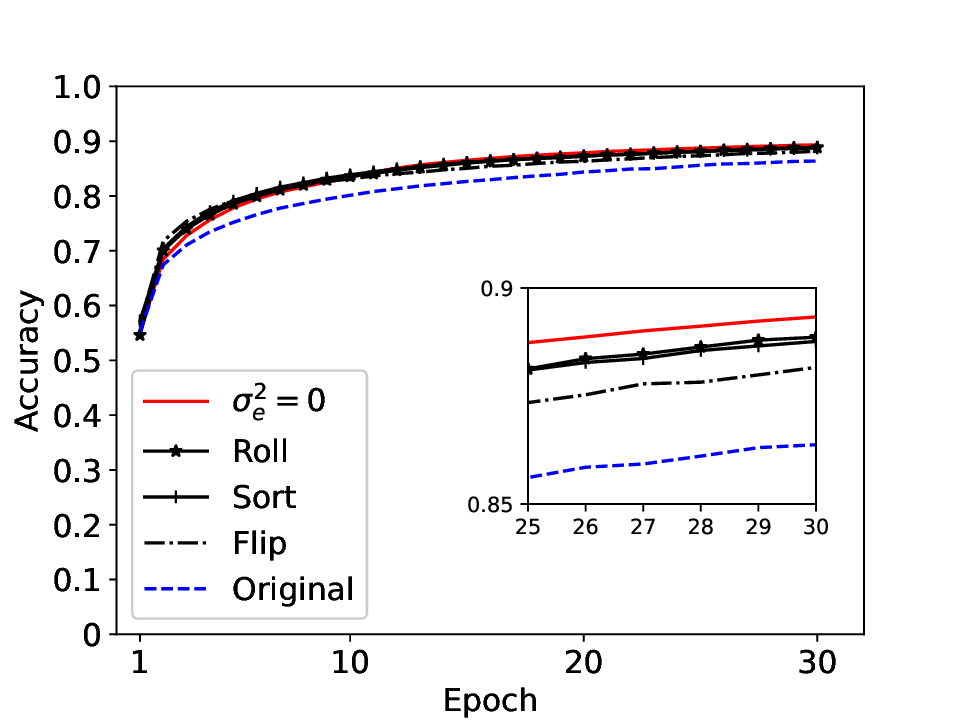}
	\end{subfigure}
	\begin{subfigure}[b]{1\columnwidth}
	\centering
	\includegraphics[width=0.9\columnwidth]{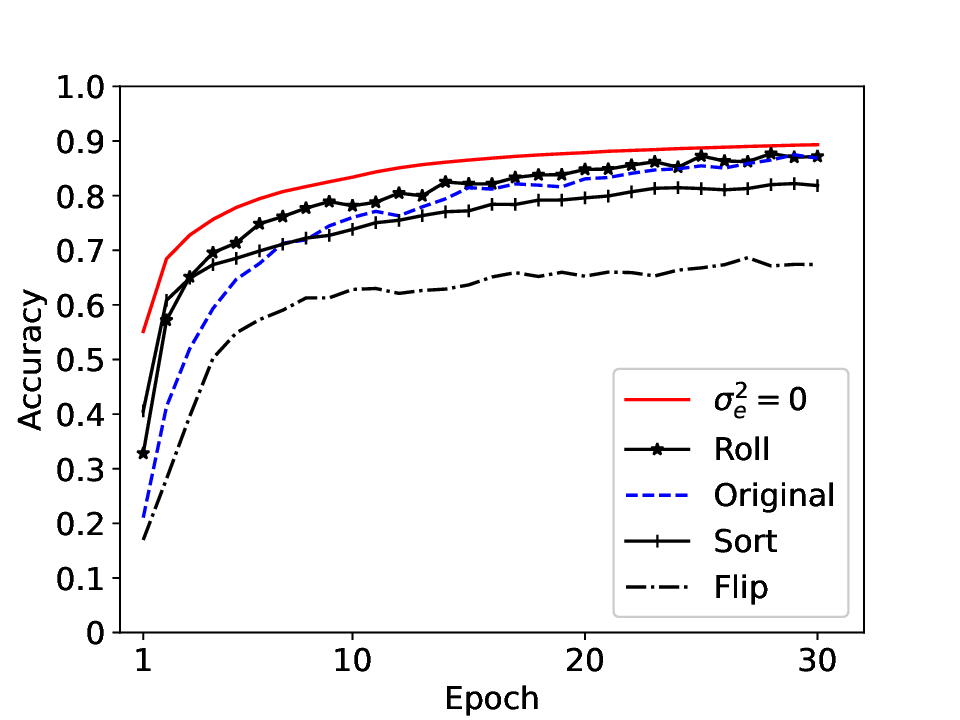}
	\end{subfigure}
	\caption{Low (top) and high (bottom) phase noise.}
	\label{fig:phase_drift_accuracy}
\end{figure}

In Figure \ref{fig:phase_drift_accuracy} the average test accuracy per epoch in the low and high phase noise scenarios is presented. The test accuracy is evaluated on the MNIST test dataset every 100 batches, meaning 12 times per epoch. This in turn is averaged over 10 i.i.d. instances of the scenario. We observe a significant effect of permuting the gradient. In the high phase noise scenario it can have both an improving and a degrading effect on learning. In the low phase noise scenario all permutations improve the learning performance.

\section{Conclusion}
We propose a scheme of permuting gradients before transmission in over-the-air computation for federated learning with phase noise. Specifically we propose the Flip, Roll and Sort permutations. Simulations show that in a scenario with high phase noise, gradient permutation can have a significant impact on the learning performance. Furthermore, using the Roll permutation appears to give the best learning performance out of the four permutations including the original permutation. In the low phase noise scenario all permutations improve the learning performance and can even cause a higher convergence rate in early epochs than in the scenario without phase noise ($\sigma_e^2=0$). The proposed permutations are not necessarily optimal, but demonstrate a significant effect of permuting the gradient before transmission. In a future work we propose studying the permutations more closely, especially the Sort permutation by computing the sorted order more frequently than once per epoch and using more batches. 

\appendix[Derivation of Variance]\label{Appendix}
Below we derive the variance given by (\ref{variance}). We start by introducing some auxiliary variables. Let $\alpha_k$ and $\rho_k$ be as follows:
\begin{equation}
	\begin{split}
		&\alpha_k \equiv \text{cos}(x), \rho_k \equiv \text{sin}(x), x \equiv -\sum_{i=0}^{d/2}e_k^{(n,i)} \sim\mathcal{N}\left(0, \frac{d}{2}\sigma^2_e\right),
	\end{split}
\end{equation}
since $\text{exp}(ix)$ has a log-normal distribution one can show that
\begin{equation}
	\begin{split}
		&\mathbb{E}[\alpha_k] = \text{exp}\left(\frac{-d\sigma^2_e}{4}\right),\mathbb{E}[\rho_k] = 0\\
		&\mathbb{E}[\alpha_k^2]  = \frac{1}{2}\left(1 + \text{exp}\left(-d\sigma^2_e\right)\right),
		\mathbb{E}[\rho_k^2] = \frac{1}{2}\left(1 - \text{exp}\left(-d\sigma^2_e\right)\right). 
	\end{split}
\end{equation}
With the first and second moments of $\alpha_k$ and $\rho_k$ we can express the second moment of the gradient estimate (\ref{estimate}). All superscripts such as $d/2$, $n$ are removed for ease of notation.

\begin{equation}
	\begin{split}
		&(\hat{g}_d)^2 = \bigg(\frac{a_d}{K}\sum_{k=0}^{K-1}(\alpha_kg_{k,d} - \rho_kg_{k,d+1})\mathbb{I}_k + \tilde{w}'\bigg)^2\\
		&=\frac{a_d^2}{K^2}\bigg(\sum_{k=0}^{K-1}\bigg((\alpha_kg_{k,d} - \rho_kg_{k,d+1})\mathbb{I}_k\bigg)^2 \\&+ \sum_{k\neq l}(\alpha_kg_{k,d} - \rho_kg_{k,d+1})\mathbb{I}_k(\alpha_lg_{l,d} - \rho_lg_{l,d+1})\mathbb{I}_l)\\
		& + \tilde{w}'\cdot\dots + (\tilde{w}')^2,
	\end{split}
\end{equation}
where $\tilde{w}' = \frac{a_d}{K}\Re(w)\sim\mathcal{N}\left(0,\sigma^2_{\tilde{w}'}\right), \sigma^2_{\tilde{w}'}=\frac{a_d^2\sigma^2_w}{2K^2}$. Then the expectation is as follows:

\begin{align}
	&\mathbb{E}[(\hat{g}_d)^2] = \frac{a_d^2}{K^2}\sum_{k=0}^{K-1}\mathbb{E}[(\alpha_kg_{k,d} - \nonumber \rho_kg_{k,d+1})^2]\mathbb{E}[\mathbb{I}_k^2] \\ \nonumber
	&+ \frac{a_d^2}{K^2}\sum_{k\neq l}\mathbb{E}[(\alpha_kg_{k,d} - \rho_kg_{k,d+1})(\alpha_lg_{l,d} - \rho_lg_{l,d+1})]\mathbb{E}[\mathbb{I}_k]^2\\\nonumber
	& + \sigma^2_{\tilde{w}'}\\ \nonumber 
	&=\frac{\text{exp}\left(\frac{t}{\sigma^2_h}+\frac{d\sigma^2_e}{2}\right)}{K^2}\sum_{k=0}^{K-1}\mathbb{E}[(\alpha_kg_{k,d} - \rho_kg_{k,d+1})^2] \\ \nonumber
	&+ \frac{\text{exp}\left(\frac{d\sigma^2_e}{2}\right)}{K^2}\sum_{k\neq l}\mathbb{E}[(\alpha_kg_{k,d} - \rho_kg_{k,d+1})(\alpha_lg_{l,d} - \rho_lg_{l,d+1})] \\ \nonumber
	&+ \sigma^2_{\tilde{w}'}\\ \nonumber
	&=\frac{\text{exp}\left(\frac{t}{\sigma^2_h}+\frac{d\sigma^2_e}{2}\right)}{2K^2}\sum_{k=0}^{K-1}\bigg(\bigg(1 + \text{exp}\left(-d\sigma^2_e\right)\bigg)g_{k,d}^2\\ \nonumber
	& + \bigg(1 - \text{exp}\left(-d\sigma^2_e\right)\bigg)g_{k,d+1}^2\bigg) + \frac{1}{K^2}\sum_{k\neq l}g_{k,d}g_{l,d} + \sigma^2_{\tilde{w}'}.\\ 
\end{align}

Finally, the variance is given by

\begin{equation}
	\begin{split}
		&\mathbb{E}[(\hat{g}_d)^2] - \mathbb{E}[\hat{g}_d]^2=\frac{\text{exp}\left(\frac{t}{\sigma^2_h}+\frac{d\sigma^2_e}{2}\right)}{2K^2}\sum_{k=0}^{K-1}\\   
		&\bigg(\bigg(1 + \text{exp}\left(-d\sigma^2_e\right)\bigg)g_{k,d}^2+ \bigg(1 - \text{exp}\left(-d\sigma^2_e\right)\bigg)g_{k,d+1}^2\bigg) \\
		&+ \frac{1}{K^2}\sum_{k\neq l}g_{k,d}g_{l,d}  - \frac{1}{K^2}\bigg(\sum_{k=0}^{K-1}g_{k,d}\bigg)^2	+ \sigma^2_{\tilde{w}'}\\
		&= \frac{\text{exp}\left(\frac{t}{\sigma^2_h}+\frac{d\sigma^2_e}{2}\right)}{2K^2}\sum_{k=0}^{K-1}\bigg(\bigg(1 + \text{exp}\left(-d\sigma^2_e\right)\\
		&-2\text{exp}\left(\frac{-t}{\sigma^2_h}-\frac{d\sigma^2_e}{2}\right)\bigg)g_{k,d}^2 + \bigg(1 - \text{exp}\left(-d\sigma^2_e\right)\bigg)g_{k,d+1}^2\bigg)+ \sigma^2_{\tilde{w}'}.
	\end{split}
\end{equation}

	\bibliographystyle{IEEEtran}
	\bibliography{ref}
\vspace{12pt}
\color{red}

\end{document}